\documentclass{iopart}
\usepackage{euscript,iopams,amssymb,amsfonts,graphicx,bm}
\usepackage{pgfplots,setstack}

\usepackage{float}
\usepackage{epsfig}
\usepackage{esint}
 \usepackage{tikz-cd}

\usepackage{bm,braket}

\eqnobysec

\newcommand{\dis}{\displaystyle}

\newcommand{\calQ}{{\mathcal Q}}
\newcommand{\calS}{{\mathcal S}}
\newcommand{\calF}{{\mathcal F}}

\newcommand{\calM}{{\mathcal M}}

\newcommand{\calN}{{\mathcal N}}

\newcommand{\R}{{\mathbb R}}

\newcommand{\X}{\mathbf{X}}
\renewcommand{\P}{\mathbb{P}}

\newcommand{\J}{\widetilde{J}}
\newcommand{\x}{\mathbf{x}}
\newcommand{\y}{\mathbf{y}}
\renewcommand{\e}{{\mathrm e}}
\newcommand{\E}{{\mathbb E}}

\newcommand{\n}{\mathbf n}
\newcommand{\z}{\mathbf z}
\newcommand{\calT}{{\mathcal T}}
\renewcommand{\P}{\mathbb P}
\newcommand{\p}{\widetilde{p}}
\renewcommand{\v}{\widetilde{v}}
\newcommand{\q}{\widetilde{q}}

\newcommand{\w}{\widetilde{w}}
\renewcommand{\S}{\widetilde{S}}

\begin{document}

\title[Diffusion with stochastic resetting screened by a semipermeable interface ]{Diffusion with stochastic resetting screened by a semipermeable interface}

\author{Paul C. Bressloff}
\address{Department of Mathematics, University of Utah 155 South 1400 East, Salt Lake City, UT 84112}

\begin{abstract} 
In this paper we consider the diffusive search for a bounded target $\Omega \in \R^d$ with its boundary $\partial \Omega$ totally absorbing. We assume that the target is surrounded by a semipermeable interface given by the closed surface $\partial \calM$ with $\Omega \subset \calM\subset \R^d$. That is, the interface totally surrounds the target and thus partially screens the diffusive search process. We also assume that the position of the diffusing particle (searcher) randomly resets
 to its initial position $\x_0$ according to a Poisson process with a resetting rate $r$. The location $\x_0$ is taken to be outside the interface, $\x_0\in \calM^c$, which means that resetting does not occur when the particle is within the interior of $\partial \calM$. (Otherwise, the particle would have to cross the interface in order to reset to $\x_0$.) Hence, the semipermeable interface also screens out the effects of resetting. We first solve the boundary value problem (BVP) for diffusion on the half-line $x\in [0,\infty)$ with an absorbing boundary at $x=0$, a semipermeable barrier at $x=L$, and stochastic resetting to $x_0>L$ for all $x>L$. We calculate the mean first passage time (MFPT) to find (be absorbed by) the target and explore its behavior as a function of the permeability $\kappa_0$ of the interface and its spatial position $L$.  In particular, we find that increasing $L$ reduces the MFPT and increases the optimal resetting rate at which the MFPT is minimized. We also find that the sensitivity of the MFPT to changes in $\kappa_0$ is a decreasing function of $L$. We then perform the analogous calculations for a three-dimensional (3D) spherically symmetric interface and target, and show that the MFPT exhibits the same qualitative behavior as the 1D case. 
Finally, we introduce a stochastic single-particle realization of the search process based on a generalization of so-called snapping out BM. The latter sews together successive rounds of reflecting Brownian motion on either side of the interface. The main challenge is establishing that the probability density generated by the snapping out BM satisfies the permeable boundary conditions at the interface. We show how this can be achieved using renewal theory.

\end{abstract}
 \maketitle
 
\section{Introduction}

A classical problem in the theory of diffusion is transport through a semipermeable interface. This type of interface arises in a wide range of natural and artificial systems. Examples at the microscopic level include artificial membranes for reverse osmosis  \cite{Li10,Rubinstein21}, lipid bilayers regulating molecular transport in biological cells \cite{Philips12,Alberts15,Bressloff21,Nik21}, and chemical and electrical gap junctions \cite{Evans02,Connors04,Good09,Bressloff16}. There are also macroscopic analogs such as animal migration in heterogeneous landscapes \cite{Beyer16,Assis19,Kenkre21}. Finally, various forms of complex and porous media are modeled in terms of multiple semipermeable interfaces and heterogeneous diffusivities \cite{Grebenkov10,Hahn12,Carr16,Aho16,Moutal19,Farago20,Alemany22}. 

At the population level, a semipermeable interface can be incorporated into the diffusion equation by imposing flux continuity across the interface, and taking the flux to be proportional to the associated  jump discontinuity in the concentration across the interface. The constant of proportionality is identified as the permeability. This permeable or leather boundary condition is a particular version of the well-known Kedem-Katchalsky (KK) equations \cite{Kedem58,Kedem62,Kargol96}, which can be derived by considering a thin membrane and using statistical thermodynamics. The KK equations also allow for discontinuities in the diffusivity and chemical potential across the interface. Although the KK equations were originally developed within the context of the transport of non-electrolytes through biological membranes, they are now used to describe all types of membranes, both biological and artificial. (See the recent collection of articles in Ref. \cite{Nik21}.) 

At the microscopic level there are two complementary methods for modeling single-particle diffusion. The first approach is to consider a random walk on a lattice, whereby diffusion is recovered in an appropriate continuum limit. In random walk models, semipermeable barriers are represented by local defects \cite{Powles92,Kenkre08,Novikov11,Kay22}. The second approach is to use stochastic differential equations (SDEs). These generate sample paths of a Brownian particle that are distributed according to a probability density satisfying the corresponding diffusion or Fokker-Planck (FP) equation. However, incorporating the microscopic analog of the permeable boundary condition is non-trivial.  A rigorous probabilistic formulation of one-dimensional (1D) BM in the presence of a semipermeable barrier has recently been introduced by Lejay \cite{Lejay16,Lejay18}, see also Refs. \cite{Aho16,Brobowski21}. This is based on so-called snapping out BM, which sews together successive rounds of partially reflecting BM that are restricted to either the left-hand or right-hand side of the barrier. (The diffusion equation for partially reflecting BM is supplemented by a Robin boundary condition at the barrier.) Suppose that the particle starts to the right of the barrier. It realizes positively reflected BM until its local time exceeds an exponential random variable with parameter $2\kappa_0$. (The local time is a Brownian functional that keeps track of the amount of time the particle spends in a neighborhood of the barrier \cite{Ito65,Majumdar05,Grebenkov06}.) It then immediately resumes either negatively or positively reflected BM with equal probability, and so on.  Using the theory of semigroups and resolvent operators, Lejay proved that snapping out BM is an exact single-particle realization of diffusion through an interface in the overdamped limit \cite{Lejay16}. (Note that SDEs in the form of underdamped Langevin equations have been used to develop efficient computational schemes for finding solutions to the FP equation in the presence of one or more semipermeable interfaces \cite{Farago18,Farago20}.)

We have recently reformulated snapping out BM in terms of a renewal equation that relates the full probability density to the probability densities of the partially reflected BMs on either side of the barrier \cite{Bressloff23a,Bressloff23b}.  (The original analysis of Lejay \cite{Lejay16} derived a corresponding backward equation.)  The renewal equation can be solved using Laplace transforms and Green's function methods, resulting in an explicit expression for the probability density of snapping out BM. We first used the renewal approach to develop a more general probabilistic model of 1D single-particle diffusion through a semipermeable barrier. This incorporated an encounter-based model of membrane absorption \cite{Grebenkov20,Grebenkov22,Bressloff22,Bressloff22a} that kills each round of partially reflected BM. In the latter case, the corresponding boundary condition at the interface involved a time-dependent permeability with memory \cite{Bressloff23b}. In subsequent work we extended the renewal theory of snapping out BM to single-particle diffusion in bounded domains and higher spatial dimensions \cite{Bressloff23b}.

In this paper we consider single-particle diffusion through a closed semipermeable interface within the context of diffusive search for a bounded target. Suppose that the target is denoted by $\Omega \subset \R^d$ with its boundary $\partial \Omega$ totally absorbing. In addition, assume that the semipermeable interface is the surface $\partial \calM$ with $\calM\subset \R^d$ and $\Omega$ a proper subset of $\calM$, see Fig. \ref{fig1} of section 2. Hence, the interface totally surrounds the target and thus partially screens the diffusive search within the unbounded domain $\R^d\backslash \Omega$. It is well-known that diffusive search for a target in an unbounded domain is characterized by an infinite mean first passage time (MFPT), irrespective of whether diffusion is recurrent ($d\leq 2$) or transient ($d>2$). One  way to render the MFPT finite is to include a stochastic resetting protocol. The simplest version is to randomly reset
 the position of the particle to some fixed position ${\bm \xi} \in \calM^c\equiv \R^d\backslash \calM$, say, according to a Poisson process with a resetting rate $r$. In recent years, stochastic resetting has emerged as an important paradigm for understanding nonequilibrium stochastic processes, with a variety of applications in optimal search problems and biophysics (see the review \cite{Evans20} and references therein.) 
  
 One of the interesting issues concerning resetting in the presence of a semipermeable barrier is how to deal with the periods when the particle is located within the interior of the interface, that is, $\X_t \in \calM \backslash \Omega$. In such situations, the particle would have to cross the interface if it were reset to ${\bm \xi}\in \calM^c$. The most natural assumption is to assume that resetting does not occur whenever $\X_t \in \calM \backslash \Omega$. That is, a closed semipermeable interface also screens out the effects of resetting. Consequently, we have a space-dependent stochastic resetting protocol. (Further examples of space-dependent resetting protocols can be found in Refs. \cite{Evans11b,Roldan17,Pinsky20}.) The main goal of this paper is to calculate the effects of screening by the semipermeable interface $\partial \calM$ on the MFPT
to find (be absorbed by) the target surface $\partial \Omega$.

The structure of the paper is as follows. In section 2, we formulate the general problem in terms of the diffusion or FP equation for the probability density of particle position. The effects of the semipermeable membrane are incorporated using a permeable boundary condition on $\partial \calM$. For simplicity we set ${\bm \xi}=\x_0\in \calM^c$, where $\x_0$ is the initial position of the particle. In section 3 we explicitly solve the boundary value problem (BVP) for diffusion on the half-line with an absorbing boundary at $x=0$, a semipermeable barrier at $x=L$, and $\xi=x_0>L$ (see Fig. \ref{fig2}). We calculate the MFPT and explore its behavior as a function of model parameters.  First, we find that the MFPT exhibits the typical unimodal dependence on the resetting rate $r$, with a unique minimum at an optimal resetting rate $r_{\rm opt}$ that depends on other model parameters. Second, as expected, the MFPT is a decreasing function of $\kappa_0$ (higher permeability). In the limit $\kappa_0\rightarrow \infty$ the barrier becomes completely permeable. However the MFPT still depends on the spatial separation $L$ due to the fact that resetting is screened out within the interval $[0,L]$. On the other hand, the MFPT diverges in the limit $\kappa_0\rightarrow 0$, since the barrier becomes impenetrable.
Third, increasing $L$ towards $x_0$ also reduces the MFPT and shifts $r_{\rm opt}$ to the right. That is, once the particle crosses to the left of the barrier it is advantageous that the region of no resetting is larger.  In addition, the effect of the semipermeable barrier relative to a fully permeable barrier increases monotonically with respect to the reset rate $r$ unless $L$ is sufficiently close to $x_0$, where the opposite occurs. In section 4, we perform the analogous calculations for a three-dimensional (3D) spherically symmetric interface and target (see Fig. \ref{fig4}), and show that the MFPT exhibits the same qualitative behavior as the 1D case. 

Finally, in section 5, we develop a single-particle realization based on a generalization of snapping out BM. The main challenge is establishing that the probability density generated by the snapping out BM satisfies the permeable boundary conditions at the interface. For the sake of illustration, we focus on the 1D case. We begin by writing down the renewal equation relating the full probability density to the probability densities of partially reflected BM on either side of the barrier. We then Laplace transform the renewal equation and use this to derive the correct boundary conditions at the interface. A subtle feature of the derivation is that it is necessary to take account of the fact that the Robin boundary condition for partially reflected BM at $x=L$ has to be modified when the particle actually starts on the boundary. (An analogous result holds in higher spatial dimensions \cite{Bressloff23b}.) The inclusion of stochastic resetting leads to a further complication, namely, when snapping out BM restarts from $x=L$ to the right of the barrier, the initial position of the resulting partially reflected BM is distinct from the reset position $x_0>L$. 

\section{Screening of a target by a semipermeable interface}

\begin{figure}[t!]
\raggedleft
\includegraphics[width=12cm]{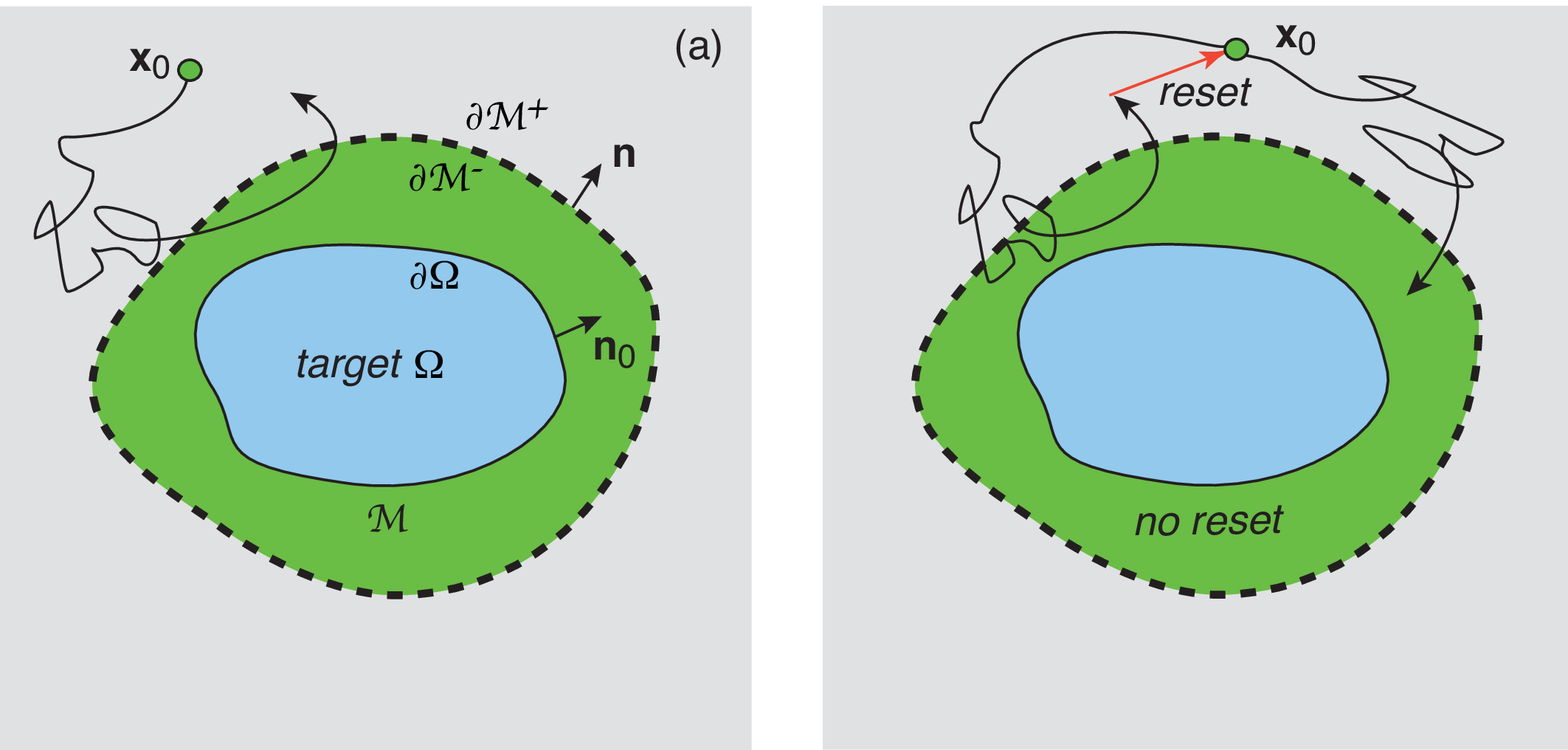}
\caption{(a) A particle diffuses in the domain $\R^d\backslash \Omega$, where $\Omega$ is an obstacle or target with a totally absorbing boundary $\partial \Omega$. The target is surrounded by a semipermeable interface $\partial \calM$ with $\calM \supset \Omega$. The shaded green (darker) region around $\Omega$ is $\calN=\calM\backslash \Omega$. The outward unit normals of $\partial \Omega$ and $\partial \calM$ are denoted by $\n_0$ and $\n$, respectively. (b) Partial stochastic resetting. Particle resets at a rate $r$ to the initial position $\x_0$ whenever it is diffusing in $\calM^c$. No resetting occurs in $\calN$.}
\label{fig1}
\end{figure}

 Consider a Brownian particle diffusing in $\R^d$ with an obstacle or target $\Omega$ whose boundary $\partial \Omega$ is totally absorbing. Furthermore, 
suppose that there exists a domain $\calM \supset \Omega$ that totally encloses the target and whose boundary $\partial \calM$ is a semipermeable interface with $\partial \calM^+$ and $\partial \calM^-$ denoting the side approached from $\calM^c$ and $\calN\equiv \calM\backslash \Omega$, respectively, see Fig. \ref{fig1}(a). Let $p_0(\x,t|\x_0)$ denote the probability density of the particle with the initial condition $\X_0=\x_0 \in \calM^c$. The density $p_0$ satisfies the diffusion equation 
\numparts
\begin{eqnarray}
\label{Rda}
\fl &\frac{\partial p_0(\x,t|\x_0)}{\partial t}=D\nabla^2p_0(\x,t|\x_0) ,\ \x \in\calN\cup \calM^c,\\
\label{Rdb}
\fl &-D\nabla p_0(\y^{\pm},t|\x_0) \cdot \n=\kappa_0[p_0(\y^-,t|\x_0)-p_0(\y^+,t|\x_0)],\quad \y^{\pm} \in \partial \calM^{\pm},\\
\fl &p_0(\x,t|\x_0)=0,\quad \x \in \partial \Omega,
\label{Rdc}
\end{eqnarray}
\endnumparts
together with the initial condition $p_0(\x,0|\x_0) =\delta(\x-\x_0)$. Here $\n$ is the unit normal directed out of $\calM$, $D$ is the diffusivity and $\kappa_0$ is the (constant) permeability. Eq. (\ref{Rdb}) is a special case of the well-known Kedem-Katchalsky (KK) boundary condition \cite{Kedem58,Kedem62,Kargol96}, which in its most general form also allows for discontinuities in the diffusivity and chemical potential across the interface. 
If the interface were impermeable ($\kappa_0=0$) then the probability of absorption would be zero for $\x_0\in \calM^c$ and unity for $\x_0\in \calN$. Moreover, in the latter case the MFPT would be finite since $\partial \calN$ is bounded. On the other hand,
if $\kappa_0>0$ then absorption can occur for all $\x_0 \in \calM^c \cup \calN$, but the MFPT for absorption is infinite since there is a nonzero probability that the  particle makes an arbitrarily large excursion away from the target. 

One way to render the MFPT finite is to include a stochastic resetting protocol \cite{Evans11a,Evans11b,Evans14}. Therefore, suppose that whenever the particle is diffusing in the domain $\calM^c$ it can reset to its initial position $\x_0\in \calM^c$ at a random sequence of times generated by a Poisson process with constant rate $r$, see Fig. \ref{fig1}(b); no resetting occurs in $\calN$. Note that the resetting protocol is space-dependent due to the fact that we exclude resetting events that involve the particle crossing the semipermeable membrane from $\calN$ to $\xi\in \calM^c$. The probability density $p_r(\x,t|\x_0)$ with resetting evolves according to
\numparts
\begin{eqnarray}
\label{pDa}
\fl &\frac{\partial p_r }{\partial t}= D\nabla^2 p_r(\x,t|\x_0)  -rp_r(\x,t|\x_0) +r S_r(\x_0,t)\delta(\x-\x_0), \, \x ,\x_0 \in  \calM^c,\\
\fl &\frac{\partial p_r }{\partial t}= D\nabla^2 p_r(\x,t|\x_0),   \, \x  \in  \calN,\\
\fl &-D\nabla p_r(\y^{\pm},t) \cdot \n=\kappa_0[p_r(\y^-,t)-p_r(\y^+,t)],\quad \y^{\pm} \in \partial \calM^{\pm},\\
\fl &p_r(\x,t|\x_0) =0 ,\ \x\in \partial \Omega.
\label{pDd}
\end{eqnarray}
\endnumparts
(All quantities with resetting are labeled by the index $r$, with the corresponding quantities without resetting obtained by setting $r=0$.) We have introduced the occupation probability
\begin{equation}
\label{pDQ}
S_r(\x_0,t)=\int_{\calM^c} p_r(\x,t|\x_0)d\x,
\end{equation}
which is the probability that the particle hasn't been absorbed in the time interval $[0,t]$ and is somewhere in the reset domain $\calM^c$, having started at $\x_0\in \calM^c$. Laplace transforming equations (\ref{pDa})-(\ref{pDd}) gives
\numparts
\begin{eqnarray}
\label{pDLTa}
\fl  & D\nabla^2 \p_r(\x,s|\x_0)  -(r +s)\p_r(\x,s|\x_0) =-\delta(\x-\x_0)[1+r \S_r(\x_0,s)],  \ \x,\x_0 \in \calM^c,  \\
\fl  & D\nabla^2 \p_r(\x,s|\x_0)  -s\p_r(\x,s|\x_0) =0,\ \x \in \calN,
\label{pDLTb} \\
\fl & -D\nabla \p_r(\y^{\pm},s) \cdot \n=\kappa_0[\p_r(\y^-,s)-\p_r(\y^+,s)],\quad \y^{\pm} \in \partial \calM^{\pm},
\label{pDLTc}\\
\fl &\p_r(\x,t|\x_0)=0, \ \x\in \partial \Omega.
\label{pDLTd}
\end{eqnarray}
\endnumparts

In this paper we are interested in determining the combined effects of the semipermeable membrane and stochastic resetting on the MFPT to find the target. Let us introduce the stopping time
\begin{equation}
\label{stop}
\calT=\inf\{t>0, \X_t \in \partial \Omega\},
\end{equation}
and the survival probability
\begin{equation}
\calQ_r(\x_0,t)=\int_{\calM^c}p_r(\x,t|\x_0)d\x,+\int_{\calN}p_r(\x,t|\x_0)d\x,
\end{equation}
The FPT density $f_r(\x_0,t)$, 
\begin{equation}
f_r(\x_0,t)dt \equiv \P[t\leq \calT\leq t+dt|\X_0=\x_0],
\end{equation}
is related to the survival probability according to $f_r(\x_0,t)=-dQ_r(\x_0,t)/dt$. Hence, the MFPT is given by
\begin{eqnarray}
\label{Tr}
\fl T_r(\x_0)\equiv \E[\calT]=\int_0^{\infty}tf(\x_0,t)dt=-\int_0^{\infty}t\frac{d\calQ_r(\x_0,t)}{dt}dt=\widetilde{\calQ}_r(\x_0,0)
\end{eqnarray}
after integration by parts, with $\widetilde{\calQ}(\x_0,s)$ the Laplace transform of the survival probability:
\begin{equation}
\widetilde{\calQ}_r(\x_0,s)=\int_0^{\infty}\e^{-st}\calQ_r(\x_0,t)dt.
\end{equation}
Similarly, higher-order moments can be expressed in terms of derivatives of $\widetilde{\calQ}_r(\x_0,s)$ with respect to $s$ at $s=0$. The survival probability can also be related to the total flux through the target using conservation of probability. That is,
\begin{equation}
\frac{d\calQ_r(\x_0,t)}{dt}=-J_r(\x_0,t)= D\int_{\partial \Omega} \nabla u_r(\x,t|\x_0)\cdot \n_0d\x,
\end{equation}
where $\n_0$ is the outward unit normal of the surface $\partial \Omega$. Laplace transforming this equation gives
\begin{equation}
\label{sur}
s\widetilde{\calQ}_r(\x_0,s)-1=-\widetilde{J}_r(\x_0,s)
\end{equation}
and, hence,
\begin{eqnarray}
\label{Tr0}
T_r(\x_0)=-\left . \frac{d\widetilde{J}_r(\x_0,s)}{ds}\right |_{s=0}
\end{eqnarray}

\section{Diffusion on the half-line}
\begin{figure}[t!]
 \raggedleft
  \includegraphics[width=8cm]{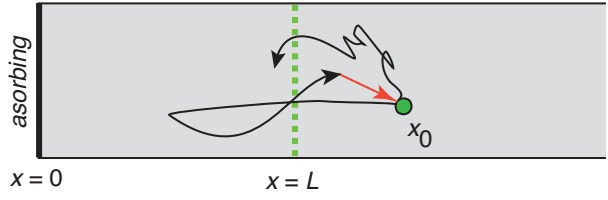}
  \caption{Semipermeable barrier at $x=L$ with an absorbing boundary at $x=0$. Resetting only occurs when $x>L$. The initial position $x_0$ is taken to be located on the right-hand side of the barrier and is identified with the reset point $\xi$. (The 2D rendition is for ease of visualization.)}
  \label{fig2}
\end{figure}

Consider the FPT problem in which a Brownian particle diffuses on the half-line $[0,\infty)$ with an absorbing boundary at $x=0$ and a semipermeable barrier at $x=L$, see Fig. \ref{fig2}. In this example, $\calN=[0,L)$, $\calM^c=(L,\infty)$, and $x_0> L$. Note that a FPT problem for diffusion through a semipermeable barrier has recently been considered in the case of a finite interval and no resetting \cite{Kay22,Bressloff23a}. Since diffusion is then unbiased, the barrier has no effect on the MFPT if the particle starts at a position from which it can reach the absorbing boundary without having to cross the semipermeable barrier. On the other hand, if the particle starts on the other side of the barrier then the MFPT does depend on $\kappa_0$ and there is a jump in the MFPT across the barrier. Since stochastic resetting introduces a bias, the MFPT is $\kappa_0$-dependent for all initial positions $x_0>0$, but there is still a discontinuity across the barrier. Given that we are interested in the screening effects of a barrier, we will focus on the case $x_0>L$.

The 1D version of equations (\ref{pDLTa})--(\ref{pDLTd}) is
  \numparts
\begin{eqnarray}
\label{1Da}
\fl&D\frac{\partial^2\widetilde{p}_r(x,s|x_0)}{\partial x^2}-(r+s) \widetilde{p}_r(x,s|x_0)=-\delta(x-x_0)[1+r\S_r(x_0,s)],\ X>L\\
\label{1Db}
\fl&D\frac{\partial^2\widetilde{p}_r(x,s|x_0)}{\partial x^2}-s \widetilde{p}_r(x,s|x_0)=0,\ 0<X<L,\\
\label{1Dc}
\fl&D\frac{\partial \widetilde{p}_r(L^+,s|x_0)}{\partial x}=D\frac{\partial \widetilde{p}_r(L^-,s|x_0)}{\partial x}=\kappa_0[ \widetilde{p}_r(L^+,s|x_0)-\p_r(L^-,s|x_0)],\\
\fl&\p_r(0,s|x_0)=0,
\label{1Dd}
\end{eqnarray}
\endnumparts
with
\begin{equation}
\label{Sr}
\S_r(x_0,s)=\int_L^{\infty} \p_r(x,s|x_0)dx.
\end{equation}
The general solution can be written in the form
\begin{equation}
\label{1Dsola}
\p_r(x,s|x_0)=A\sinh(\sqrt{s/D}x),\ 0\leq x<L,
\end{equation}
\begin{eqnarray}
\label{1Dsolb}
\fl \widetilde{p}_r(x,s|x_0)&=\left \{B\e^{\sqrt{(r+s)/D}[x-L]}+C\e^{-\sqrt{(r+s)/D}[x-L]}\right \}\e^{-\sqrt{(r+s)/D}(x_0-L)} 
\end{eqnarray}
for $L<x<x_0$, with the solution for $x>x_0$ obtained by switching $x$ and $x_0$ in (\ref{1Dsolb}). This follows from continuity of the solution across $x_0$. However, the Dirac delta function on the right-hand side of equation (\ref{1Da}) means that there is a discontinuity in the first derivative of $\p_r$ across $x_0$:
\begin{equation}
 D\frac{\partial \widetilde{p}_r(x_0^+,s|x_0)}{\partial x}-D\frac{\partial \widetilde{p}_r(L^-,s|x_0)}{\partial x}=-(1+r\S_r(x_0,s)).
 \end{equation}
 It follows that
 \begin{equation}
 \label{B}
 B=\frac{1+r\S_r(x_0,s)}{2\sqrt{[r+s]D}}.
 \end{equation}
 
 The coefficients $A$ and $C$ are determined by imposing the permeable boundary conditions (\ref{1Dc}). The latter yield the pair of equations
 \numparts
  \begin{eqnarray}
  \label{cona}
\fl & [B-C]\e^{-\sqrt{(r+s)/D}(x_0-L)}=\sqrt{\frac{s}{r+s}}A \cosh(\sqrt{s/D}L),\\
 \fl &[B+C]\e^{-\sqrt{(r+s)/D}(x_0-L)}=A\left \{\frac{\sqrt{sD} }{\kappa_0}\cosh(\sqrt{s/D}L)+\sinh(\sqrt{s/D}L)\right \}.
 \label{conb}
 \end{eqnarray}
 \endnumparts
 Adding these equations gives
 \begin{equation}
 \fl 2B\e^{-\sqrt{(r+s)/D}(x_0-L)}=A\left \{\left [\frac{\sqrt{sD} }{\kappa_0}+\sqrt{\frac{s}{r+s}}\right ]\cosh(\sqrt{s/D}L)+\sinh(\sqrt{s/D}L)\right\},
 \end{equation}
 and substituting for $B$ using equation (\ref{B}) we have $A=A_r(x_0,s)$ with
 \begin{eqnarray}
 \label{A}
 \fl &A_r(x_0,s)\\
 \fl &=\frac{[1+r\S_r(x_0,s)]\e^{-\sqrt{(r+s)/D}(x_0-L)} }{\left [\sqrt{(r+s)D} \sqrt{sD} /\kappa_0+\sqrt{sD}\right ]\cosh(\sqrt{s/D}L)+\sqrt{(r+s)D}\sinh(\sqrt{s/D}L)}.\nonumber 
 \end{eqnarray}
 Similarly, the coefficient $C$ is obtained by subtracting equations (\ref{cona}) and (\ref{conb}).
 It remains to determine the Laplace transform of the occupation probability, $\S_r(x_0,s)$. One way to proceed would be to substitute the solution $\p_r(x,s|x_0)$, $L\leq x<\infty$ into equation (\ref{Sr}) and to evaluate the resulting integral. However, since we are ultimately interested in calculating the MFPT, we only require $\lim_{s\rightarrow 0} \S_r(x_0,s)$. The latter can be determined in terms of the flux through the semipermeable barrier.
 
 The first step is to calculate the flux through the absorbing boundary at $x=0$:
 \begin{equation}
 \J_{r,0}(x_0,s)=D\frac{\partial \p_r(0,s|x_0)}{\partial x}=\sqrt{sD}A_r(x_0,s).
 \end{equation}
 It follows that
 \begin{equation}
 \label{JL}
 \lim_{s\rightarrow 0}\J_{r,0}(x_0,s)=\frac{[1+r\lim_{s\rightarrow 0}\S_r(x_0,s)]\e^{-\sqrt{r/D}(x_0-L)} }{\sqrt{rD}/\kappa_0 +1 +\sqrt{r/D}L}.
 \end{equation}
 Next, using similar arguments to the derivation of equation (\ref{sur}) and exploiting flux continuity across the $x=L$, we have
 \begin{eqnarray}
\fl s \S_r(x_0,s)-1&=-\J_{r,L^{\pm}}(x_0,s)\equiv -D\frac{\partial \widetilde{p}_r(L^{\pm},s|x_0)}{\partial x}\nonumber \\
\fl & =-\sqrt{sD}A_r(x_0,s)\cosh(\sqrt{s/D}L)=- \J_{r,0}(x_0,s)\cosh(\sqrt{sD}L).
\label{JLL}
 \end{eqnarray}
 Since the probability of absorption approaches unity in the large time limit (for $0<r<\infty$), we have
 \begin{equation}
 \lim_{t\rightarrow \infty} S_r(x_0,t)=\lim_{s\rightarrow 0}s\S_r(x_0,s)=0.
 \end{equation}
 Therefore, taking the limit $s\rightarrow 0$ in equation (\ref{JLL}) implies that $ \lim_{s\rightarrow 0}\J_{r,0}(x_0,s)=1$ so that (\ref{JL}) reduces to the condition
  \begin{equation}
\frac{[1+r\lim_{s\rightarrow 0}\S_r(x_0,s)]\e^{-\sqrt{r/D}(x_0-L)} }{\sqrt{rD}/\kappa_0 +1 +\sqrt{r/D}L}=1.
 \end{equation} 
 Finally, rearranging this equation, we have
 \begin{equation}
 \label{Sr0}
\lim_{s\rightarrow 0}\S_r(x_0,s)=\frac{ \sqrt{rD}/\kappa_0 +1 +\sqrt{r/D}L}{r}\e^{\sqrt{r/D}(x_0-L)}-\frac{1}{r}.
 \end{equation} 
 
 We can now calculate the MFPT using the 1D version of equation (\ref{Tr}). The Laplace transformed survival probability is
 \begin{eqnarray}
 \widetilde{\calQ}_r(x_0,s)&=\int_0^{\infty} \p_r(x,s|x_0)dx=\int_0^L\p_r(x,s|x_0)dx+\S_r(x_0,s)\nonumber \\
 &=A_r(x_0,s) \sqrt{D/s}(\cosh(\sqrt{s/D}L)-1)+\S_r(x_0,s).
 \end{eqnarray}
 Hence, using the Taylor expansion $\cosh(\sqrt{s/D}L)=1+sL^2/2D+O(s^2)$, we have
 \begin{eqnarray}
 \fl T_r(x_0)&=\lim_{s\rightarrow 0} \widetilde{\calQ}_r(x_0,s)=\lim_{s\rightarrow 0} \left \{ \sqrt{sD}A_r(x_0,s) [L^2/2D +O(s)]+\S_r(x_0,s\right \}\nonumber\\
\fl & =\frac{ \sqrt{rD}/\kappa_0 +1 +\sqrt{r/D}L}{r}\e^{\sqrt{r/D}(x_0-L)}-\frac{1}{r}+\frac{L^2}{2D}.
\label{Tr1D}
 \end{eqnarray}
A number of general features emerge from this result:

\begin{enumerate}
\item In the limit $\kappa_0 \rightarrow 0$ the barrier becomes impermeable, and $T_r(x_0)\rightarrow \infty$. This follows from the assumption $x_0>L$ so that the particle can never reach the absorbing boundary.

\item In the limit $\kappa_0 \rightarrow \infty$ the barrier is completely permeable. The MFPT becomes
 \begin{eqnarray}
 T_r(x_0) =\frac{ 1 +\sqrt{r/D}L}{r}\e^{\sqrt{(r+s)/D}(x_0-L)}-\frac{1}{r}+\frac{L^2}{2D}.
 \end{eqnarray}
 The dependence on $L$ reflects the fact that no resetting occurs in the interval $[0,L]$. That is, the target at $x=0$ is still screened form resetting. In the additional limit $L\rightarrow 0$, we recover the standard formula for the MFPT for diffusion with resetting on the half-line \cite{Evans11a,Evans11b}:
  \begin{eqnarray}
 T_r(x_0) =\frac{ \e^{\sqrt{(r+s)/D}(x_0-L)}-1}{r} .
 \end{eqnarray}
 
 \item The MFPT diverges in the limit $r\rightarrow 0$ since the MFPT for pure diffusion on the half-line is infinite even though it is recurrent (absorption occurs with unit probability). The MFPT also diverges in the limit $r\rightarrow \infty$, since the particle returns to its initial position so frequently it cannot reach the origin. As shown below, we find that $T_r(x_0)$ exhibits the familiar unimodal dependence on $r$, with a unique minimum at an optimal resetting rate $r_{\rm opt}$ that depends on other model parameters.
\end{enumerate}
 
 \begin{figure}[b!]
 \raggedleft
  \includegraphics[width=13cm]{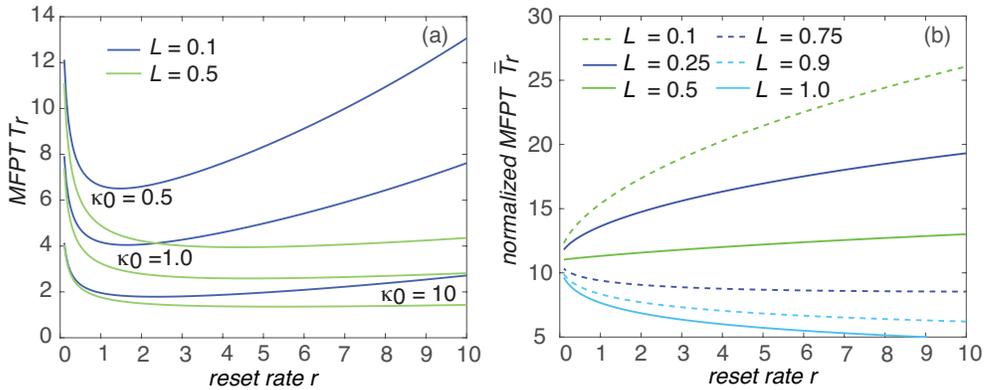}
  \caption{Diffusion on the half-line. (a) Plots of the MFPT $T_r(x_0)$ as a function of the reset rate $r$ for various values of the barrier distance $L$ and the permeability $\kappa_0$. (b) Corresponding plots of the normalized MFPT $\overline{T}_r(x_0)$ for $\kappa_0=0.1$. We also set $D=1$ and $x_0=1$.}
  \label{fig3}
\end{figure}

In Fig. \ref{fig3}(a) we show sample plots of the MFPT $T_r(x_0)$ as a function of the reset rate $r$ for various values of the barrier distance $L$ and the permeability $\kappa_0$. We observe typical unimodal curves, each with a minimum at an optimal reset rate $r_{\rm opt}$. As expected, the curves shift downwards as $\kappa_0$ increases (higher permeability). Increasing $L$ towards $x_0$ also reduces the MFPT and shifts $r_{\rm opt}$ to the right. That is, once the particle crosses to the left of the barrier it is advantageous that the region of no resetting is larger. In order to separate out more clearly the dependence of the MFPT on  $\kappa_0$ and $L$ from $r$, we introduce the normalized MFPT
\begin{equation}
\overline{T}_r(x_0)=\frac{T_r(x_0)}{\lim_{\kappa_0 \rightarrow \infty}T_r(x_0)}.
\end{equation}
In Fig. \ref{fig3}(b) we plot $\overline{T}_r$ as a function of $r$ for fixed $\kappa_0=0.1$ and different values of $L$. One result that emerges from this figure is as follows: the effect of the semipermeable barrier relative to a fully permeable barrier increases monotonically with respect to the reset rate $r$ unless $L$ is sufficiently close to $x_0$, where the opposite occurs. A similar result holds for other values of $\kappa_0$.

\section{Spherically symmetric semipermeable interface and target}

\begin{figure}[t!]
  \raggedleft
  \includegraphics[width=7cm]{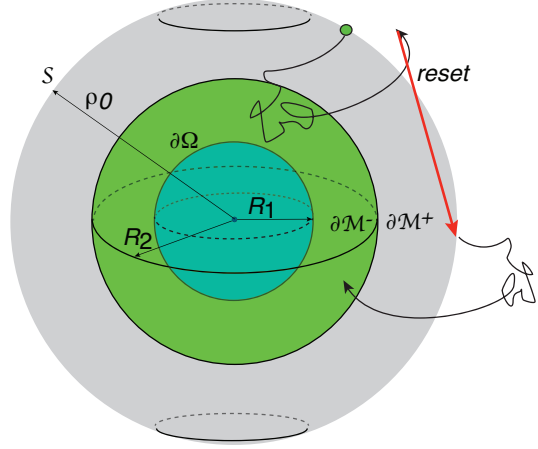}
  \caption{Diffusion through a spherically symmetric semipermeable interface $\partial \calM \subset \R^3$ of radius $R_2$, which is concentric with a totally  absorbing target of radius $R_1$, $R_1<R_2$. The initial condition and reset point are uniformally distributed on a sphere $\calS$ of radius $\rho_0>R_2$}
  \label{fig4}
\end{figure}

As our second example configuration, suppose that $\Omega =\{\x\in \R^3\,|\, 0\leq  |\x| <R_1\}$ and $\calM =\{\x\in \R^3\,|\, 0\leq  |\x| <R_2\}$ with $R_1<R_2$. The corresponding absorbing and semi-permeable surfaces are $\partial \Omega=\{\x\in \R^d\,|\,  |\x| =R_1\}$ and $\partial \calM =\{\x\in \R^d\,|\,  |\x| =R_2\}$, see Fig. \ref{fig4}.
Following \cite{Redner01}, we assume that the particle starts from and resets to a random point on the initial surface $\calS =\{\x\in \R^3\,|\,  |\x| =\rho_0\}$. We can then exploit spherical symmetry by setting $\p_r=\p_r(\rho,s|\rho_0)$,
where $\rho= |{\bf x}|$.  Equations (\ref{pDLTa})-- (\ref{pDLTd}) reduce to the form
\numparts
\begin{eqnarray}
\fl &D\frac{\partial^2\p_r(\rho,s|\rho_0)}{\partial \rho^2} + \frac{2D}{\rho}\frac{\partial \p_r(\rho,s|\rho_0)}{\partial \rho}-(r+s)\p_r(\rho,s|\rho_0) \nonumber \\
\fl &\quad =-\frac{\delta(\rho-\rho_0)}{4\pi\rho_0^2}[1+\S_r(\rho_0,s)] ,\  R_2 <\rho ,
\label{spha}\\
\label{sphb}
\fl &D\frac{\partial^2\p_r(\rho,s|\rho_0)}{\partial \rho} + \frac{2D}{\rho}\frac{\partial \p_r(\rho,s|\rho_0)}{\partial \rho}-s\p_r(\rho,s|\rho_0) =0 ,\  R_1<\rho<R_2,\\
\fl   &D\frac{\partial \p_r(R_2^+,s|\rho_0)}{\partial \rho}=D\frac{\partial \p_r(R_2^-,s|\rho_0)}{\partial \rho}=\kappa_0 [\p_r(R_2^+,s|\rho_0) -\p_r(R_2^-,s|\rho_0)],\\
\fl &\p_r(R_1,s|\rho_0)=0.
\label{sphd}
\end{eqnarray}
\endnumparts
The occupation probability is
\begin{equation}
\S_r(\rho_0,s)=4\pi \int_{R_2}^{\infty}\p_r(\rho,s|\rho_0)\rho^2d\rho.
\end{equation}

A well known trick for solving the spherically symmetric diffusion equation in 3D is to perform the change of variable $\p_r(\rho,s|\rho_0)=v_r(\rho,s|\rho_0)/\rho$. Substituting into equations (\ref{spha})--(\ref{sphd}) yields an effective 1D diffusion equation:
\numparts
\begin{eqnarray}
\label{1spha}
\fl &D\frac{\partial^2\v_r(\rho,s|\rho_0)}{\partial \rho^2} -(r+s)\v_r(\rho,s|\rho_0) =-\frac{\delta(\rho-\rho_0)}{4\pi\rho_0} [1+r\S_r(\rho_0,s)] , R_2 <\rho ,\\
\label{1sphb}
\fl &D\frac{\partial^2\v_r(\rho,s|\rho_0)}{\partial \rho} -s\v_r(\rho,s|\rho_0) =0 ,\  R_1<\rho<R_2,\\
\label{1sphc}
\fl   &D\frac{\partial \v_r(R_2^+,s|\rho_0)}{\partial \rho}-\frac{D}{R_2}\v_r(R_2^+,s|\rho_0)= \kappa_0 [\v_r(R_2^+,s|\rho_0)-\v_r(R_2^-,s|\rho_0)],\\
\fl   &D\frac{\partial \v_r(R_2^-,s|\rho_0)}{\partial \rho}-\frac{D}{R_2}\v_r(R_2^-,s|\rho_0)= \kappa_0 [\v_r(R_2^+,s|\rho_0)-\v_r(R_2^-,s|\rho_0)],
\label{1sphd}\\
\fl &\v_r(R_1,s|\rho_0)=0.
\label{1sphe}
\end{eqnarray}
\endnumparts
Comparison with equations (\ref{1Da})--(\ref{1Dd}) shows that $\v_r$ is given by a modified version of the solutions (\ref{1Dsola}) and (\ref{1Dsolb}): 
\begin{equation}
\label{3Dsola}
\v_r(\rho,s|\rho_0)=\overline{A} \sinh(\sqrt{s/D}[\rho-R_1]),\ R_1\leq \rho<R_2,
\end{equation}
\begin{eqnarray}
\label{3Dsolb}
\fl \v_r(\rho,s|\rho_0)&=\left \{\overline{B}\e^{\sqrt{(r+s)/D}[\rho-R_2]}+\overline{C}\e^{-\sqrt{(r+s)/D}[\rho-R_2]}\right \}\e^{-\sqrt{(r+s)/D}(\rho_0-R_2)}  
\end{eqnarray}
for $R_2<\rho < \rho_0$ with
\begin{equation}
\label{spB}
\overline{B}=\frac{1+r\S_r(\rho_0,s)}{8\pi \rho_0 \sqrt{(r+s)D}}.
\end{equation}
Similarly, imposing the permeable boundary conditions (\ref{1sphc}) and (\ref{1sphd}) yields the analog of equations (\ref{cona}) and (\ref{conb}):
 \numparts
  \begin{eqnarray}
\fl  &\sqrt{[r+s]D}[\overline{B}-\overline{C}]\e^{-\sqrt{(r+s)/D}(\rho_0-R_2)}=\left [\kappa_0+\frac{D}{R_2}\right ][\overline{B}+\overline{C}]\e^{-\sqrt{(r+s)/D}(\rho_0-R_2)}\nonumber \\ 
\fl &\hspace{6cm}-\kappa_0\overline{A} \sinh(\sqrt{s/D}\Delta R),\\
 \fl &\sqrt{sD}\, \overline{A}\cosh(\sqrt{s/D}\Delta R)=\kappa_0[\overline{B}+\overline{C}]\e^{-\sqrt{(r+s)/D}(\rho_0-R_2)}\nonumber \\ 
\fl &\hspace{6cm}-\left [\kappa_0-\frac{D}{R_2}\right ]\overline{A} \sinh(\sqrt{s/D}\Delta R),
 \end{eqnarray}
 \endnumparts
 where $\Delta R=R_2-R_1$.
 Rearranging these equations we have
  \numparts
  \begin{eqnarray}
  \label{spcona}
\fl& [\overline{B}+\overline{C}]\e^{-\sqrt{(r+s)/D}(\rho_0-R_2)}\\
\fl&= \frac{\overline{A}}{\kappa_0} \bigg \{\sqrt{sD}\cosh(\sqrt{s/D}\Delta R)+\left [\kappa_0-\frac{D}{R_2}\right ]\sinh(\sqrt{s/D}\Delta R)\bigg \},\nonumber\\
\fl &[\overline{B}-\overline{C}]\e^{-\sqrt{(r+s)/D}(\rho_0-R_2)}\nonumber \\
 \fl&=\overline{A} \frac{\kappa_0+D/R_2}{\kappa_0\sqrt{(r+s)D}}\bigg \{\sqrt{sD}\cosh(\sqrt{s/D}\Delta R)+\left [\kappa_0-\frac{D}{R_2}\right ]\sinh(\sqrt{s/D}\Delta R)\bigg \},\nonumber \\ 
\fl &\hspace{6cm}-\frac{\kappa_0}{\sqrt{(r+s)D}}\overline{A} \sinh(\sqrt{s/D}\Delta R).
 \label{spconb}
 \end{eqnarray}
 \endnumparts
Adding these equations gives
 \begin{eqnarray}
 \fl &2\overline{B}\e^{-\sqrt{(r+s)/D}(\rho_0-R_2)}\nonumber \\
 \fl & \quad =\frac{\overline{A}}{\kappa_0} \left [1+\frac{\kappa_0+D/R_2}{\sqrt{(r+s)D}}\right ]\bigg \{\sqrt{sD}\cosh(\sqrt{s/D}\Delta R)+\left [\kappa_0-\frac{D}{R_2}\right ]\sinh(\sqrt{s/D}\Delta R)\bigg \}\nonumber \\
 \fl &\hspace{6cm}-\frac{\kappa_0}{\sqrt{(r+s)D}}\overline{A} \sinh(\sqrt{s/D}\Delta R),
 \end{eqnarray}
 and substituting for $\overline{B}$ using equation (\ref{spB}) we have $\overline{A}=\overline{A}_r(\rho_0,s)$ with
 \begin{eqnarray}
 \label{barA}
 \fl &\overline{A}_r(\rho_0,s)=\frac{1}{4\pi \rho_0}\frac{[1+r\S_r(\rho_0,s)]\e^{-\sqrt{(r+s)/D}(\rho_0-R_2)} }{\Gamma_1(s)\sqrt{sD}\cosh(\sqrt{s/D}\Delta R)+\Gamma_2(s)\sinh(\sqrt{s/D}\Delta R)},
 \end{eqnarray}
 with
 \begin{eqnarray}
\fl \Gamma_1(s)= [\sqrt{(r+s)D} +\kappa_0+D/R_2]\frac{1}{\kappa_0},\quad \Gamma_2(s)=\Gamma_1(s)\left [\kappa_0-\frac{D}{R_2}\right ] -\kappa_0.
 \end{eqnarray}

 We will calculate $\lim_{s\rightarrow 0} \S_r(\rho_0,s)$ along similar lines to the 1D case. First, the flux through the absorbing boundary at $\rho=R_1$ is
 \begin{eqnarray}
 \J_{r,1}(\rho_0,s)&=\left . 4\pi R_1^2D\frac{\partial \p_r(\rho,s|\rho_0)}{\partial \rho}\right |_{\rho=R_1}\nonumber \\
 &=4\pi R_1 \left \{D\left . \frac{\partial \v_r(\rho,s|\rho_0)}{\partial \rho}\right |_{\rho=R_1}-\frac{D}{R_1}\v_r(R_1,s|\rho_0)\right \}\nonumber \\
 &=4\pi R_1 \sqrt{sD} \, \overline{A}_r(\rho_0,s).
 \end{eqnarray}
 It follows that
 \begin{eqnarray}
 \label{spJL}
\fl  &\lim_{s\rightarrow 0}\J_{r,0}(\rho_0,s)=\frac{R_1}{\rho_0}\frac{[1+r\lim_{s\rightarrow 0}\S_r(x_0,s)]\e^{-\sqrt{r/D}(\rho_0-R_2)} }{ \calF_r(\kappa_0,R_1,R_2)},
 \end{eqnarray}
 where
 \begin{eqnarray}
\fl  \calF_r(\kappa_0,R_1,R_2)&= [\sqrt{rD}/\kappa_0+1+D/(R_2\kappa_0)][1+(\kappa_0-D/R_2)\Delta R/D)]-\kappa_0\Delta R/D\nonumber \\
\fl &=\frac{R_1}{R_2}\left (\frac{\sqrt{rD}}{\kappa_0}+\frac{D}{\kappa_0 R_2}\right )+1+\sqrt{\frac{r}{D}}\Delta R.
\label{Fr}
 \end{eqnarray}
 Next, from conservation of probability
 \begin{eqnarray}
\fl & s \S_r(\rho_0,s)-1=-\J_{r,R_2^{\pm}}(\rho_0,s)\equiv -\left . 4\pi R_2^2 D\frac{\partial \widetilde{v}_r(\rho,s|x_0)/\rho}{\partial \rho}\right |_{\rho=R_2^{\pm}}\nonumber \\
\fl &=-4\pi R_2 \left \{D\left . \frac{\partial \v_r(\rho,s|\rho_0)}{\partial \rho}\right |_{\rho=R_2}-\frac{D}{R_2}\v_r(R_2,s|\rho_0)\right \}\nonumber \\
\fl & =-4\pi R_2\left \{ \sqrt{sD}\, \cosh(\sqrt{s/D}\Delta R)-\frac{D}{R_2}\sinh(\sqrt{s/D}\Delta R) \right \}\overline{A}_r(\rho_0,s).
\label{SSr}
 \end{eqnarray}
 Since the probability of absorption approaches unity in the large time limit (for $0<r<\infty$), we have
 \begin{equation}
 \lim_{t\rightarrow \infty} S_r(x_0,t)=\lim_{s\rightarrow 0}s\S_r(x_0,s)=0.
 \end{equation}
 Therefore, taking the limit $s\rightarrow 0$ in equation (\ref{SSr}) implies that 
 \begin{eqnarray}
 4\pi R_1 \lim_{s\rightarrow 0} \sqrt{sD}\overline{A}_r(\rho_0,s)\equiv \lim_{s\rightarrow 0}\J_{r,0}(\rho_0,s)=1.
 \end{eqnarray}
Hence, from equation (\ref{spJL}) we have
 \begin{eqnarray}
  \label{spSr0}
\fl & \lim_{s\rightarrow 0}\S_r(\rho_0,s) =\frac{\rho_0}{R_1} \frac{\calF_r(\kappa_0,R_1,R_2)\e^{\sqrt{r/D}(\rho_0-R_2)}}{r}-\frac{1}{r}.
 \end{eqnarray} 
 
 \begin{figure}[b!]
 \raggedleft
  \includegraphics[width=13cm]{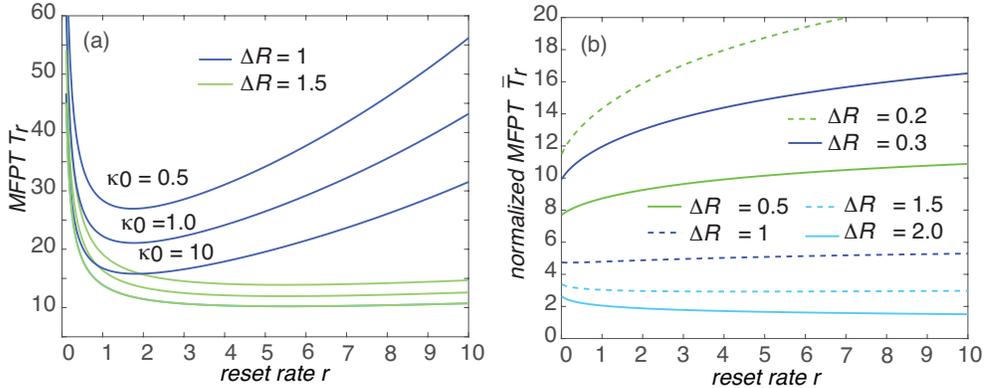}
  \caption{3D diffusion with spherical symmetry. (a) Plots of the MFPT $T_r(\rho_0)$ as a function of the reset rate $r$ for various values of the radial separation $\Delta R$ (for fixed $R_1$) and the permeability $\kappa_0$. (b) Corresponding plots of the normalized MFPT $\overline{T}_r(x_0)$ for a fixed permeability $\kappa_0=0.1$. We also set $D=1$, $\rho_0=3$ and $R_1=1$.}
  \label{fig5}
\end{figure}

\begin{figure}[t!]
 \raggedleft
  \includegraphics[width=13cm]{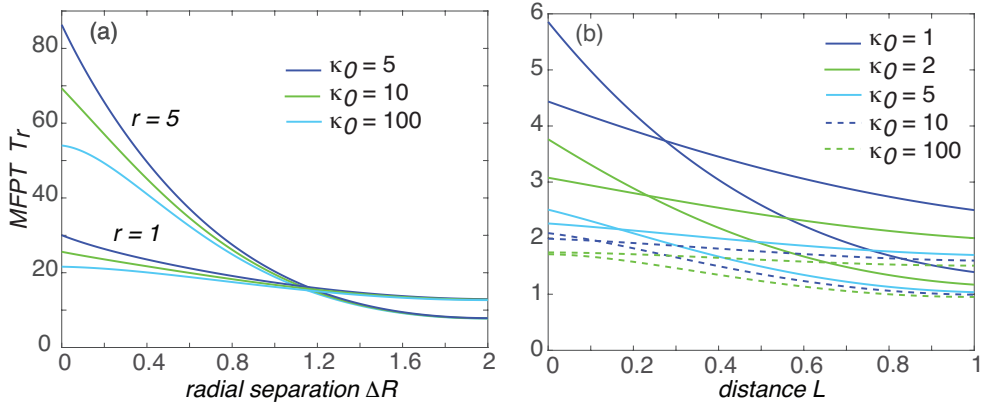}
  \caption{(a) 3D diffusion with spherical symmetry. Plots of the normalized MFPT $\overline{T}_r(x_0)$ as a function of the reset rate $r$ for various values of the interface radius $R_2$ and fixed permeability $\kappa_0=0.1$. We also set $D=1$, $\rho_0=3$ and $R_1=1$. Note that the curves do not exactly coincide when  $R_2=R_0$. (b) Corresponding plots for the half-line with $D=1$ and $x_0=1$. For each value of $\kappa_0$ the curve that decreases more quickly (slowly) corresponds to $r=5$ ($r=1$).}
  \label{fig6}
\end{figure}

 We can now calculate the MFPT using the 3D version of equation (\ref{Tr}) with spherical symmetry. The Laplace transformed survival probability is
 \begin{eqnarray}
\fl &\widetilde{\calQ}_r(x_0,s)=4\pi \int_{R_1}^{R_2} \p_r(\rho,s|\rho_0)\rho^2 d\rho+\S_r(\rho_0,s)\nonumber \\
 \fl &=4\pi \overline{A}_r(x_0,s) \int_{R_1}^{R_2} \sinh(\sqrt{s/D}[\rho-R_1)\rho d\rho +\S_r(x_0,s)\nonumber \\
 \fl &=4\pi \overline{A}_r(x_0,s) \sqrt{\frac{D}{s}}\left (R_2\cosh(\sqrt{s/D}\Delta R)- \sqrt{\frac{D}{s}}\sinh(\sqrt{s/D}\Delta R)-R_1 \right )+\S_r(x_0,s)\nonumber \\
 \fl &=4\pi \sqrt{sD}\overline{A}_r(x_0,s) \left \{\frac{R_2(\Delta R)^2}{2D}-\frac{(\Delta R)^3}{6D}+O(s)\right \}+\S_r(x_0,s).
 \end{eqnarray}
 Hence,  
 \begin{eqnarray}
 \fl T_r(x_0)&=\lim_{s\rightarrow 0} \widetilde{\calQ}_r(x_0,s) \nonumber \\
\fl &=\frac{\rho_0}{R_1} \frac{\calF_r(\kappa_0,R_1,R_2)\e^{\sqrt{r/D}(\rho_0-R_2)}}{r}-\frac{1}{r}+\frac{(\Delta R)^2}{2D}\frac{2R_2+R_1}{3R_1}.
\label{Tr3D}
 \end{eqnarray}
It follows from equation (\ref{Fr}) that $\calF_r\rightarrow \infty$ as $\kappa_0\rightarrow 0$, reflecting the fact that the interface at $\rho=R_2$ is now impenetrable. On the other hand, in the limit $\kappa_0 \rightarrow \infty$ the MFPT converges to
\begin{equation}
\fl T_r(\rho_0)=\frac{\rho_0}{R_1} \frac{\left [1+\sqrt{\frac{r}{D}}\Delta R\right ]\e^{\sqrt{r/D}(\rho_0-R_2)}}{r}-\frac{1}{r}+\frac{(\Delta R)^2}{2D}\frac{2R_2+R_1}{3R_1}.
\end{equation}
Analogous to the 1D case, the MFPT depends on the radius $R_2$ of the totally penetrable interface, since resetting only occurs for $\rho\geq R_2$. If we also set $R_2=R_1$ so that $\Delta R=0$ then we recover the standard result \cite{Evans14}
\begin{equation}
T_r(\rho_0)=\frac{\rho_0}{R_1} \frac{ \e^{\sqrt{r/D}(\rho_0-R_1)}}{r}-\frac{1}{r}.
\end{equation}
Finally, note that the MFPT diverges when $R_1\rightarrow 0$, since the target disappears. 

For fixed target radius $R_1$, the qualitative behavior of the MFPT is similar to the 1D case under the mappings $\Delta R \rightarrow L$ and $\rho_0\rightarrow x_0$. This is illustrated in Figs. \ref{fig5}(a,b), which are the 3D analogs of Figs. \ref{fig3}(a,b). In Fig. \ref{fig6}(a), we plot the MFPT as a function of the radial separation $\Delta R=R_2-R_1$ for fixed $R_1$ and different choices of $\kappa_0$ and $r$. This not only shows that the MFPT is a decreasing function of $\Delta R$ but also indicates that there is a significant reduction in the sensitivity to the permeability $\kappa_0$ as $R_2\rightarrow \rho_0$. The latter effect is much weaker in the case of the half-line, as illustrated in Fig. \ref{fig6}(b).

\section{Single particle realization using snapping out Brownian motion}

\begin{figure}[t!]
 \raggedleft
  \includegraphics[width=13cm]{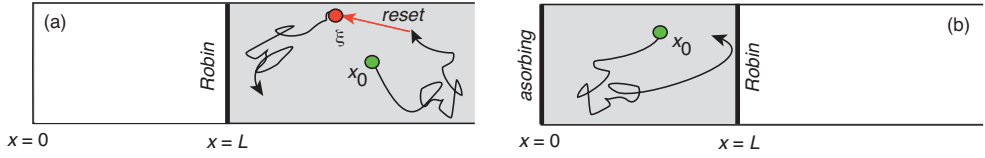}
  \caption{Decomposition of snapping out BM on the half-line into two partially reflected BMs. (a) Partially reflected BM in $[L,\infty)$ with stochastic resetting to a point $\xi>L$. (b) Partially reflected BM in $[0,L]$ supplemented by a totally absorbing boundary at $x=0$ and no resetting. }
  \label{fig7}
\end{figure}

So far we have focused on the forward diffusion or Fokker-Planck equation for the probability density of particle position in the presence of a semipermeable membrane and stochastic resetting. A nontrivial problem is determining the underlying stochastic differential equation (SDE) that realizes sample paths of the given stochastic process. As we highlighted in the introduction, the appropriate SDE framework is snapping out BM \cite{Lejay16,Bressloff23a,Bressloff23b}. 
 In this section we show how to construct snapping out BM in the presence of an absorbing target and stochastic resetting. For the sake of illustration, we focus on the example of diffusion on the half-line, see Fig. \ref{fig2}. We first analyze partially reflecting BMs in the domains $[0,L]$ and $[L,\infty)$, see Fig. \ref{fig7}, and then show how to sew them together to generate a generalized snapping out BM.
 
 \subsection{Partially reflected BM in $[0,L]$} 
 
 Consider BM in the interval $[0,L]$ with $x=0$ totally absorbing and $x=L$ partially reflecting. Let $X_t\in [0,L]$ denote the position of the Brownian particle at time $t$ and introduce the Brownian local time
\begin{equation}
\label{loc}
\ell_t=\lim_{\epsilon\rightarrow 0} \frac{D}{\epsilon} \int_0^tH(\epsilon-|X_{\tau}-L|)d\tau,
\end{equation}
where $H$ is the Heaviside function. Note that $\ell_t$, which has units of length due to the additional factor of $D$, determines the amount of time that the Brownian particle spends in the neighborhood of $x=L$ over the interval $[0,t]$. It can be shown that $\ell_t$ exists and is a nondecreasing, continuous function of $t$. The partially reflecting boundary condition at $x=L$ is implemented by introducing the stopping time
\begin{equation}
\calT_L=\inf\{t>0: \ell_t>\widehat{\ell}\},\quad \P[\widehat{\ell}>\ell]\equiv \Psi(\ell)=\e^{-\kappa_0\ell/D}.
\end{equation}
That is the stochastic process is killed when the local time exceeds a random exponentially distributed threshold \cite{Ito65,Freidlin85,Papanicolaou90,Milshtein95,Borodin96,Grebenkov06}. Note that $\calT_L=\infty$ if the particle reaches $x=0$ before $\ell_t$ crosses $\widehat{\ell}$. The latter event occurs at the stopping time
\begin{equation}
\calT_0=\inf\{t>0: X_t=0\}.
\end{equation}
with $\calT_0=\infty$ if the particle is first absorbed at $x=L$.
Prior to absorption at one of the ends, the position $X_t$ evolves according to the SDE
\begin{equation}
dX_t=\sqrt{2D}dW_t-d\ell_t,
\end{equation}
where $W_t$ is a Wiener process and
\begin{equation}
d\ell_t=\delta(X_t-L)dt.
\end{equation}
It can be shown that the corresponding probability density for particle position,
\begin{equation}
q(x,t|x_0)dx=\P[x\leq X_t<x+dx, t<\calT|X_0=x_0],
\end{equation}
satisfies the FP equation with a Robin boundary condition at $x=L$:
  \numparts
\label{Robin2}
\begin{eqnarray}
\frac{\partial q(x,t|x_0)}{\partial t}&=D\frac{\partial^2 q(x,t|x_0)}{\partial x^2}, \quad 0<x<L,\\
D\partial_xq(L,t|x_0)&=-\kappa_0 q(L,t|x_0),\ q(0,t|x_0)=0,
\end{eqnarray}
\endnumparts
and $q(x,0|x_0)=\delta(x-x_0)$.

 In Laplace space we have
  \numparts
\begin{eqnarray}
\label{RobinLTa}
&D\frac{\partial^2\widetilde{q}(x,s|x_0)}{\partial x^2}-s \widetilde{q}(x,ts|x_0)=-\delta(x-x_0),\\
&D\partial_x\widetilde{q}(L,s|x_0)=-\kappa_0 \widetilde{q}(L,s|x_0),\ \q(L,s|x_0)=0,
\label{RobinLTb}
\end{eqnarray}
\endnumparts
with $0<x,x_0<L$.
We can identify $\widetilde{q}(x,s|x_0)$ as a Green's function of the modified Helmholtz equation on $[0,L]$. It has the explicit form
\begin{eqnarray}
 \widetilde{q}(x,s|x_0)= \left \{ \begin{array}{cc} A \widetilde{q}_<(x,s)\widetilde{q}_>(x_0,s), & 0\leq x\leq x_0\\ & \\
 A\widetilde{q}_>(x,s)\widetilde{q}_<(x_0,s), & x_0\leq x\leq L\end{array}
 \right .
 \label{solq}
 \end{eqnarray}
 with
 \numparts
 \begin{equation}
\widetilde{q}_<(x,s)= \sinh(\sqrt{s/D}x),
\end{equation}
and
\begin{eqnarray}
\widetilde{q}_>(x,s)&= \frac{1}{2}\left [\e^{\sqrt{s/D}[L-x]}+\frac{\sqrt{sD}-\kappa_0}{\sqrt{sD}+\kappa_0}\e^{-\sqrt{s/D}[L-x]}\right ]   \\
&=\frac{\kappa_0\sinh(\sqrt{s/D}[L-x])+\sqrt{sD}\cosh(\sqrt{s/D}[L-x])}{\sqrt{sD}+\kappa_0}.\nonumber
\end{eqnarray}
\endnumparts
Matching the discontinuity in the first derivative determines the coefficient $A$:
  with
 \begin{equation}
  A=A(\kappa_0,s)\equiv \frac{(\sqrt{sD}+\kappa_0)/\sqrt{sD}} {\sqrt{sD}\cosh(\sqrt{sD}L)+\kappa_0\sinh(\sqrt{s/D}L)}.
  \end{equation}
 In particular, note that
 \begin{equation}
 \widetilde{q}(x,s|L)= \frac{\sinh(\sqrt{s/D}x)}{\sqrt{sD}\cosh(\sqrt{s/D}L)+\kappa_0\sinh(\sqrt{s/D}L)},
 \end{equation}
and
 \begin{equation}
 \label{kol}
 D\partial_x\widetilde{q}(L,s|L)=-\kappa_0\widetilde{q}(L,s|L)+1.
 \end{equation}
Note that the boundary condition (\ref{kol}) when $x_0=L$ is a modified version of the Robin boundary condition when $x_0<L$. 

 \subsection{Partially reflected BM in $[L,\infty)$} 
 
Note that this particular example has also been analyzed in Refs. \cite{Evans13,Bressloff22b}. Let $w_r(x,t|x_0,\xi)$ denote the probability density of partially reflected BM in $[L,\infty)$ with stochastic resetting to a given point $\xi>L$ at a Poisson rate $r$. (For the moment we do not identify $\xi$ with the initial position $x_0$.) Then
\numparts
\begin{eqnarray}
\label{wDa}
\fl &\frac{\partial w_r }{\partial t}= D\frac{\partial^2 w_r(x,t|x_0,\xi)}{\partial x^2}  -rw_r(x,t|x_0,\xi) +r S_r(x_0,\xi,t)\delta(x-\xi), \, x ,x_0 >L\\
\fl &D\frac{\partial w_r(L,t|x_0,\xi)}{\partial x}=\kappa_0w_r(L,t|x_0,\xi).
\label{wDb}
\end{eqnarray}
\endnumparts
Note that $S_r(x_0,\xi,t)$ is the survival probability for the given partially reflected BM,
\begin{equation}
S_r(x_0,\xi,t)=\int_L^{\infty}w_r(x,s|x_0,\xi)dx.
\end{equation} Laplace transforming equations (\ref{wDa}) and (\ref{wDb}), we have
\numparts
\begin{eqnarray}
\label{wDaLT}
\fl &D\frac{\partial^2 \w_r(x,s|x_0,\xi)}{\partial x^2}  -(r+s)\w_r(x,s|x_0,\xi) =-\delta(x-x_0)-r \S_r(x_0,\xi,s)\delta(x-\xi), \, \\
\fl &D\frac{\partial \w_r(L,s|x_0,\xi)}{\partial x}=\kappa_0\w_r(L,s|x_0,\xi).
\label{wDbLT}
\end{eqnarray}
\endnumparts
 The solution for $r=0$ is given by
  \begin{eqnarray}
 \widetilde{w}_0(x,s|x_0)=  \left \{ \begin{array}{cc}\dis  \frac{1}{\sqrt{sD}} \widetilde{w}_<(x,s)\widetilde{w}_>(x_0,s), & L\leq x\leq x_0,\\ & \\ \dis
\frac{1}{\sqrt{sD}} \widetilde{w}_>(x,s)\widetilde{w}_<(x_0,s), & x_0\leq x\leq \infty,\end{array}
 \right .
 \label{solw}
 \end{eqnarray}
 with
 \numparts
 \begin{eqnarray}
\widetilde{w}_<(x,s)&= \frac{1}{2}\left [\e^{\sqrt{s/D}[x-L]}+\frac{\sqrt{sD}-\kappa_0}{\sqrt{sD}+\kappa_0}\e^{-\sqrt{s/D}[x-L]}\right ]   \\
&=\frac{\sqrt{sD}\cosh(\sqrt{s/D}[x-L])+\kappa_0\sinh(\sqrt{s/D}[x-L])}{\sqrt{sD}+\kappa_0},\nonumber
\end{eqnarray}
and
\begin{equation}
\widetilde{w}_>(x,s)=\e^{-\sqrt{s/D}(x-L)}.
\end{equation}
\endnumparts
In particular, 
 \begin{equation}
\fl  \widetilde{w}_0(x,s|L)=\frac{\e^{-\sqrt{s/D}(x-L)}}{\sqrt{sD} +\kappa_0 },\ D\partial_x\widetilde{w}_0(L,s|L)=\kappa_0\widetilde{w_0}(L,s|L)-1.
 \label{wkol}
 \end{equation}
 Note that it is necessary to modify the Robin boundary condition when the particle starts at the barrier. This plays a crucial role in establishing the equivalence of the probability density for snapping out BM with the solution of the corresponding diffusion equation, see also Ref. \cite{Bressloff23b}.

Using the fact that $\w_0(x,s|x_0)$ is the Green's function for the partially reflecting BM without resetting, i.e set $r=0$ in equations (\ref{wDaLT}) and (\ref{wDbLT}), it follows that
  \begin{eqnarray}
\label{w1D}
  \w_r(x, s|x_0,\xi) =  \w_0(x,r+s|x_0) + r\S_r(x_0,\xi,s)\w_0(x,r+s|\xi)  .
\end{eqnarray}
In addition,
 \begin{eqnarray}
\fl  \S_r(x_0,\xi,s)&=\int_{L}^{\infty}\w_r(\,s|x_0,\xi)dx=\S_0(x_0,r+s) +r\S_r(x_0,\xi,s)\S_0(\xi,r+s),
\label{wQQr}
 \end{eqnarray}
 where $\S_0$ is the Laplace transform of the survival probability without resetting:
  \begin{eqnarray}
 \S_0(x_0,s)
 &=\frac{\sqrt{sD}+\kappa_0(1-\e^{-\sqrt{s/D}(x_0-L)})}{s(\sqrt{sD}+\kappa_0)}.
 \end{eqnarray}
  Rearranging equation (\ref{wQQr}) thus determines the survival probability with resetting in terms of the corresponding probability without resetting:
 \begin{equation}
 \label{wSr}
 \S_r(x_0,\xi,s)=\frac{\S_0(x_0,r+s)}{1-r\S_0(\xi,r+s)},
 \end{equation}
 and, hence,
 \begin{eqnarray}
\label{wpr}
  \w_r(x, s|x_0,\xi) = \w_0(x,r+s|x_0)+\frac{\w_0(x,r+s| \xi)r\S_0(x_0,r+s)}{1- r\S_0(\xi,r+s)}.
\end{eqnarray}
Note that if $\xi>L$ then
 \begin{eqnarray}
 \fl  D\partial_x\w_r(L, s|L,\xi) &=D \partial_x\w_0(L,r+s|L)+D\frac{\partial_x\w_0(L,r+s| \xi)r\S_0(L,r+s)}{1- r\S_0(\xi,r+s)}\nonumber \\
 \fl &=\kappa_0\w_0(L,r+s|L_0)-1+\frac{\kappa_0 \w_0(L,r+s| \xi)r\S_0(L,r+s)}{1- r\S_0(x_0,r+s)}\nonumber \\
 \fl &=\kappa_0\w_r(L, s|L,\xi)-1.
 \label{BCw1}
\end{eqnarray}
Explicitly differentiating between the initial position and the reset position plays a crucial role in constructing the renewal equation for snapping out BM.

\subsection{Renewal equation for snapping out BM}

We now construct a generalized version of 1D snapping out BM introduced by Lejay \cite{Lejay16} by following our renewal equation approach \cite{Bressloff23a,Bressloff23b}. This exploits the fact that snapping out BM satisfies the strong Markov property.\footnote{Recall that a continuous stochastic process $\{X_t\, \ t\geq 0\}$ is said to have the Markov property if the conditional probability distribution of future states of the process (conditional on both past and present states) depends only upon the present state, not on the sequence of events that preceded it. That is, for all $t'>t$ we have $\P[X_{t'}\leq x|X_{s},s\leq t]=\P[X_{t'}\leq x|X_{t}]$.
 The {strong Markov property} is similar to the Markov property, except that the ``present'' is defined in terms of a stopping time.} Suppose that the particle starts at $x_0>L$ and realizes positively reflected BM with resetting, see Fig. \ref{fig7}(a), until its local time $\ell_t$ at $x=L^+$ is greater than an independent exponential random variable $\widehat{\ell}$ of parameter $\kappa_0$. The process immediately restarts as a new reflected BM with probability 1/2 in either $[0,L^-]$ or $[L^+,\infty)$ and a new local time. Again the reflected BM is stopped when the local time exceeds a new exponential random variable etc. Let $p_r(x,t|x_0)$ denote the probability density of snapping out BM. Following along the lines of Refs. \cite{Bressloff23a,Bressloff23b}, $p_r$ satisfies a last renewal equation of the form
\numparts
 \begin{eqnarray}
 \label{renewala}
\fl   & p_r(x,t|x_0)=\frac{\kappa_0}{2} \int_0^t q_r(x,\tau|L)[p_r(L^+,t-\tau |x_0)+p_r(L^-,t-\tau )]d\tau  
    \end{eqnarray}
    for $x\in [0,L^-]$ and
     \begin{eqnarray}
    \fl & p_r(x,t|x_0)=w_r(x,t|x_0,x_0) +\frac{\kappa_0}{2} \int_0^t w_r(x,\tau|L,x_0)[p_r(L^+,t-\tau )+p_r(L^-,t-\tau )]d\tau  \nonumber \\
\fl  \label{renewalb}
    \end{eqnarray}
    \endnumparts
    for $x\in [L^+,\infty)$ and $\xi=x_0$. The first term on the right-hand side of equation (\ref{renewalb}) represents all sample trajectories that have never been absorbed by the barrier at $x=L^{+}$ up to time $t$. (Since $x_0 >L$ there is no analogous term in equation (\ref{renewala}).) The integral terms in equations (\ref{renewala}) and (\ref{renewala}) sum over all trajectories that were last absorbed (stopped) at time $t-\tau$ in the partially reflected BM state to the right or left of the barrier, and then switched with probability 1/2 to the appropriate side in order to reach $x$ at time $t$. Since the particle is not absorbed over the interval $(t-\tau,t]$, the probability of reaching $x$ is $q_r(x,\tau|0)$ and $w_r(x,t|x_0,x_0) $, respectively. Finally, the probability that the last stopping event occurred in the interval $(t-\tau,t-\tau+d\tau)$ irrespective of previous events is $\kappa_0 d\tau$. 
    
We now establish that  above snapping out BM is the single-particle realization of the stochastic process underlying the diffusion equation (\ref{1Da})--(\ref{1Dd}). (Note that a crucial element of the proof is that we have to distinguish between the initial position and the reset point, due to the presence of the term $w_r(x,\tau|L,x_0)$ in equation (\ref{renewalb}) where $X_0=L$ and $\xi=x_0$.) Clearly $p_r$ satisfies the diffusion equation in the bulk and the absorbing boundary condition at $x=0$, so we will focus on the boundary conditions at the semipermeable barrier. Laplace transforming the renewal equations (\ref{renewala}) and (\ref{renewalb}) with respect to time $t$ and using the convolution theorem gives
 \numparts
 \begin{eqnarray}
  \label{renewal2a}
\fl  \p_r(x,s|x_0) &= \frac{\kappa_0}{2} \q_r(x,s|L)\Sigma_{p}(x_0,s),\, x\in [0,L^-], \\
\fl  \p_r(x,s|x_0) &= \w_r(x,s|x_0,x_0)+ \frac{\kappa_0}{2} \w_r(x,s|L,x_0)\Sigma_{p}(x_0,s),\,x\in [L^+,\infty).
    \label{renewal2b}
 \end{eqnarray}
 \endnumparts
 where
 \begin{equation}
 \Sigma_{p}(x_0,s)=\p_r(L^-,s|x_0 )+\p_r(L^+,s|x_0 ) .
 \end{equation}
 Setting $x=L^+$ and $x=L^-$ in equations (\ref{renewal2a}) and (\ref{renewal2b}), respectively, summing the results and rearranging shows that 
  \begin{eqnarray}
  \label{Lam0}
\Sigma_{p}(x_0,s) = \frac{ \w_r(L,s|x_0,x_0)}{1- \kappa_0[\q(L,s|L)+\w_r(L,s|L,x_0)]/2} .
 \end{eqnarray}
 Next, differentiating equations (\ref{renewal2a}) and (\ref{renewal2b}) with respect to $x$ and setting $x=L^{\pm}$ gives
  \numparts
  \label{pho}
  \begin{eqnarray}
\fl D\partial_x \p_r(L^-,s|x_0)&=\frac{\kappa_0}{2}D\partial_x\q(L,s|L)\Sigma_{p}(x_0,s),\\
\fl D\partial_x \p_r(L^+,s|x_0)&=D\partial_x\w_r(L,s|x_0,x_0)+\frac{\kappa_0}{2}
 D\partial_x\w_r(L,s|L,x_0)\Sigma_{\rho}(x_0,s) .
  \end{eqnarray}
  \endnumparts
We now apply the Robin boundary conditions (\ref{RobinLTb}) and (\ref{wDbLT}) together with the modified Robin boundary conditions at $x=L$, namely, equations (\ref{kol}) and (\ref{BCw1}). This yields
   \numparts
  \begin{eqnarray}
    \label{pho2a}
\fl D\partial_x \p_r(L^-,s|x_0)&=-\frac{\kappa_0}{2}[\kappa_0\q(L,s|L)-1]\Sigma_{p}(x_0,s) ,\\
\fl D\partial_x \p_r(L^+,s|x_0)&=\kappa_0\w_r(L,s|x_0,x_0)+\frac{\kappa_0}{2}
 [\kappa_0\w_r(L,s|L,x_0)-1]\Sigma_{p}(x_0,s) .
   \label{pho2b}
  \end{eqnarray}
  \endnumparts
Subtracting equations (\ref{pho2a}) and (\ref{pho2b}) implies that
   \begin{eqnarray}
\fl&D[\partial_x \p_r(L^-,s|x_0)-\partial_x \p_r(L^+,s|x_0)] \nonumber\\
\fl&\quad =\kappa_0\Sigma_p(x_0,s) - \frac{\kappa_0^2 }{2}[\q(L,s|L)+\w_r(L,s|L,x_0)]\Sigma_{p}(x_0,s)-\kappa_0\w_r(L,s|x_0,x_0)\nonumber\\
\fl &\quad =0.
 \label{dev1}
  \end{eqnarray}
Similarly, adding equations (\ref{pho2a}) and (\ref{pho2b}) gives
  \begin{eqnarray}
 \fl  2 D\partial_x \p_r(L^{\pm},s|x_0)&=\kappa_0\w_r(L,s|x_0,x_0)+\frac{\kappa_0^2 }{2}[\q(L,s|L)+\w_r(L,s|L,x_0)]\Sigma_{p}(x_0,s)\nonumber \\
 \fl &=\kappa_0 [\p_r(L^-,s|x_0 )+\p_r(L^+,s|x_0 ) ].
\label{dev2}
\end{eqnarray}
 
 Equations (\ref{dev1}) and (\ref{dev2}) establish that the density $\p_r(x,s|x_0)$ of the generalized snapping out BM satisfies the diffusion equation (\ref{1Da})--(\ref{1Dd}) under the mapping $\kappa_0\rightarrow \kappa_0/2$. Hence, the snapping out BM $X_t$  is the single-particle realization of the stochastic process whose probability density evolves according to the diffusion equation in the presence of a semipermeable membrane at $x=L$ with permeability $\kappa_0/2$, a totally absorbing boundary at $x=0$ and stochastic resetting to $x_0$, $x_0>L$,  at a Poisson rate $r$. It can be checked that the solution of the renewal equation recovers the results of section 3 after the scaling $\kappa_0\rightarrow \kappa_0/2$..

\begin{figure}[b!]
\raggedleft
\includegraphics[width=12cm]{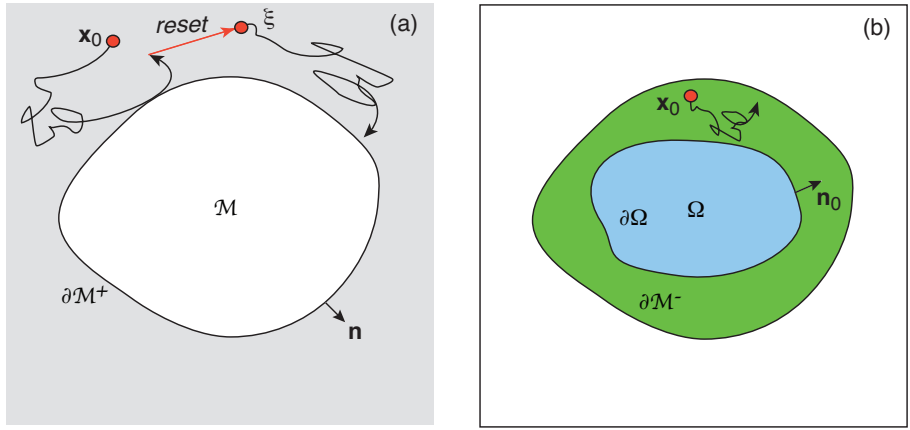}
\caption{Decomposition of snapping out BM into two partially reflected BMs. (a) Partially reflected BM in $\calM^c$ with stochastic resetting to ${\bm \xi}$. (b) Partially reflected BM in $\calN=\calM\backslash \Omega$ supplemented by a totally absorbing boundary condition on $\partial \Omega$.}
\label{fig8}
\end{figure}

\subsection{Snapping out BM in $R^d$}

An analogous construction holds for the higher-dimensional case shown in Fig. \ref{fig1}. Here we only sketch the basic steps. First we decompose snapping out BM into two complementary partial reflecting BMs. The first occurs in $\calM^c$ with $\partial \calM^+$ taken to be partially reflecting, see Fig. \ref{fig8}(a). As in the previous example, we initially distinguish between the initial position $\x_0$ and the reset point ${\bm \xi}$. The second partially restricted BM is restricted to the domain $\calN=\calM\backslash \Omega$ with $\partial \calM^-$ partially reflecting and $\partial \Omega$ totally absorbing, see Fig. \ref{fig8}(b). The probability densities in the two domains are denoted by $w_r(\x,t|\x_0,{\bm \xi})$ and $q(\x,s|\x_0)$, respectively. They evolve according to the BVPs
\numparts
\begin{eqnarray}
\label{w3Da}
\fl &\frac{\partial w_r }{\partial t}= D\nabla^2 w_r(\x,t|\x_0,{\bm \xi})  -rw_r(\x,t|\x_0,{\bm \xi}) +r S_r(\x_0,{\bm \xi},t)\delta(\x-{\bm \xi}), \, \x   \in  \calM^c,\\
\fl &D\nabla p_r(\y,t) \cdot \n=\kappa_0p_r(\y^-,t), \quad \y \in \partial \calM,
\label{w3Dd}
\end{eqnarray}
\endnumparts
and
\numparts
\begin{eqnarray}
\label{q3Da}
 &\frac{\partial q }{\partial t}= D\nabla^2 q(\x,t|\x_0) , \, \x   \in  \calN,\\
&-D\nabla q(\y,t|\x_0) \cdot \n=\kappa_0q(\y,t|\x_0),\quad \y \in \partial \calM,\\
 &q(\x,t|\x_0) =0 ,\ \x\in \partial \Omega.
\label{q3Dd}
\end{eqnarray}
\endnumparts
The higher-dimensional renewal equation takes the form
   \numparts
\begin{eqnarray}
\label{drenewala}
 \fl & p_r(\x,t|\x_0)=w_r(\x,t|\x_0,{\bm \xi})\\
 \fl &\hspace{2cm}+\frac{\kappa_0}{2}\int_0^t \left \{ \int_{\partial \calM}w_r(\x,\tau|\z,{\bm \xi})[p_r(\z^+,t-\tau|\x_0) +p_r(\z^-,t-\tau|\x_0 )]d\z\right \}d\tau ,\nonumber
 \end{eqnarray}
 for $\x\in {\calM^c}$ and
\begin{eqnarray}
\fl   p_r(\x,t|\x_0)&=\frac{\kappa_0}{2}\int_0^t \left \{ \int_{\partial \calM}q(\x,\tau|\z)[p_r(\z^+,t-\tau|\x_0 ) +p_r(\z^-,t-\tau|\x_0 )]d\z\right \}d\tau 
   \label{drenewalb}
   \end{eqnarray}
   for $\x\in   {\calN }$.
 \endnumparts 
 The proof that the solution of the renewal equation $p_r(\x,t|\x_0)$ satisfies the BVP given by equations (\ref{pDa})--(\ref{pDd}) when ${\bm \xi}=\x_0$ and $\kappa_0\rightarrow \kappa_0/2$ proceeds along analogous lines to the 1D case. Again, it is necessary to use the fact that the corresponding Robin boundary conditions for the densities $q(\x,t|\x_0)$ and $w_r(\x,t|\x_0,\xi)$ need to be modified when $\x_0\in \partial \calM$. The details of the modification can be found in \cite{Bressloff23b}. 

\section{Discussion} In this paper we explored the screening effects of a semipermeable interface on the diffusive search for a single absorbing target. In the presence of stochastic resetting to point(s) exterior to the interface, we also assumed that the interface shields the target from the effects of resetting. We proceeded by solving the BVPs for two simple geometric configurations:  (i) 1D diffusion on the half-line with an absorbing boundary (target) at $x=0$ and a semipermeable barrier at $x=L$; (ii) 3D diffusion with a spherically symmetric target and interface. The interfacial boundary conditions maintained
 flux continuity across the interface, with the flux proportional to an associated  jump discontinuity in the concentration. The constant of proportionality was identified as the permeability. In both examples, we calculated the MFPT to be absorbed by the target and determined its dependence on the permeability $\kappa_0$, resetting rate $r$, and the distance between the target and interface. Finally, we showed the equivalence between the solution of the BVP for diffusion with the solution to a renewal equation whose probability density is generated by sample paths of a generalized form of snapping out BM. Although we calculated the MFPTs in this paper by solving the diffusion equation rather than the corresponding renewal equation, the latter has at least two potential advantages. First, since snapping out BM generates sample paths of single-particle diffusion through semipermeable interfaces, it can be used to develop numerical schemes for generating solutions to the corresponding BVP see also \cite{Farago18,Farago20}. Second, the renewal equation provides a framework for developing more general probabilistic models along the lines considered in Refs. \cite{Bressloff23a,Bressloff23b}.
 
 One natural extension of the above theory is to consider the effects of interfacial screening on multiple targets. Such a set up is suggestive of a generic problem in cell biology, in which the semipermeable interface represents the lipid membrane of a cell and the multiple targets correspond to subcellular compartments. One could also consider a higher level model in which multiple cells with multiple targets compete for molecular components.

\end{document}